\documentstyle[12pt]{article}

\newcommand{\eq}{\begin{equation}}
\newcommand{\eqn}{\begin{displaymath}}
\newcommand{\en}{\end{equation}}
\newcommand{\enn}{\end{displaymath}}

\def\pnot{\mbox{${\not{\hbox{\kern-3.0pt$p$}}}$}}
\def\qnot{\mbox{${\not{\hbox{\kern-2.0pt$q$}}}$}}
\def\enot{\mbox{${\not{\hbox{\kern-2.0pt$e$}}}$}}
\def\knot{\mbox{${\not{\hbox{\kern-2.0pt$k$}}}$}}
\def\la{\mathrel{\mathpalette\fun <}}

\def\fun#1#2{\lower3.6pt\vbox{\baselineskip0pt\lineskip.9pt\ialign
{$\mathsurround=0pt#1\hfil##\hfil$\crcr#2\crcr\sim\crcr}}}

\begin{document}
\begin{titlepage}


\hskip 12cm \vbox{\hbox{DFUC}\hbox{July 1993}}
\vskip 0.6cm
\centerline{\bf RADIATIVE  CORRECTIONS}
\centerline{\bf TO QUARK-QUARK-REGGEON VERTEX IN QCD$^{\ast}$}
\vskip 1.4cm
\centerline{  V.Fadin$^{a\dagger}$, R.Fiore$^{\ddagger}$, A.Quartarolo}
\vskip .5cm
\centerline{\sl  Dipartimento di Fisica, Universit\`a della Calabria}
\centerline{\sl Istituto Nazionale di Fisica Nucleare, Gruppo collegato di
Cosenza}
\centerline{\sl Arcavacata di Rende, I-87036 Cosenza, Italy}
\vskip 3cm
\begin{abstract}
One loop corrections to the coupling of the reggeized gluon to quarks are
calculated in QCD. Combining this result with the known corrections to
the gluon-gluon-reggeon vertex, we check the self-consistency of the
representation
of the amplitudes with gluon quantum numbers and negative signature in the $t$-
channel in terms of Regge pole contribution.

\end{abstract}
\vskip .5cm
\hrule
\vskip.3cm
\noindent

\noindent
$^{\ast}${\it Work supported in part by the Ministero dell'Universit\`a e della
Ricerca Scientifica e Tecnologica}
\vfill
$^{a}${{\it Permanent address}: Budker Institute for Nuclear Physics and
Novosibirsk State University, Novosibirsk, Russia}

$ \begin{array}{ll}
^{\dagger}\mbox{{\it email address:}} &
 \mbox{FADIN@INP.NSK.SU}\\
\end{array}
$

$ \begin{array}{ll}
^{\ddagger}\mbox{{\it email address:}} &
 \mbox{40330::FIORE}\\
 & \mbox{FIORE@COSENZA.INFN.IT}
\end{array}
$
\vfill
\end{titlepage}
\eject
{\bf 1. Introduction}

Since a long time it has been found~\cite{FKL} that, in the leading logarithmic
approximation (LLA) for the Regge region, the total cross section
${\sigma}^{LLA}_{tot}$ in the non-Abelian SU({N}) gauge theories grows at large
c.m.s. energies ${\sqrt{s}}$:
\eq
{\sigma}^{LLA}_{tot}\sim {s^{\omega_{0}}\over {\sqrt{\ln{s}}}}~,
\label{z1}\en
where
\eq
{\omega_{0}} = {g^2\over {\pi}^2}N{\ln{2}}~.
\label{z2}\en
Therefore the Froissart bound ${\sigma_{tot}}<c~{\ln^2{s}}$ is violated in
the LLA. The reason for this is the violation of the $s$-channel unitarity
constraints for scattering amplitudes in the LLA.
\vskip.3cm
The behaviour (\ref{z1}) of the total cross section is determined by the
position of the most right singularity in the complex momentum plane in the
solution of the integral equation for $t$-channel partial waves with vacuum
quantum numbers~\cite{FKL}. In order to find out the region in which the LLA
can be applied, radiative corrections to the equation's kernel must be
calculated. The calculation of these corrections was started
by L.N. Lipatov and one of the authors (V.F.) in ref.~\cite{LF}, where the
calculation program was presented. The program makes a strong use of the
gluon reggeization proven in LLA~\cite{FKL}. As a necessary step in this
program one needs to calculate one loop corrections to the
particle-particle-reggeon (PPR) vertices. Here, the reggeon is
the reggeized gluon
and its trajectory in the LLA is given by
\eqn
j(t) = 1+\omega(t)~,\enn
\eq
 \omega(t) = {g^2t\over 16{\pi}^3}{N}\int{d^2k\over{{\bar k}^2({\bar q}-
{\bar k})^2}}~,~~~~~~t = -{\bar q}^2~.
\label{z3}\en
The infrared divergence in the gluon trajectory (\ref{z3}) is cancelled
by the divergences in real gluon emission, so that the integral
equation for the $t$-channel partial waves with vacuum quantum numbers
{}~\cite{FKL} is free of singularities. In order to remove the infrared
divergences at intermediate steps we use the dimensional regularization of
Feynman integrals:
\eq
{d^2k\over {{(2\pi)}^2}}~~\rightarrow ~~{d^{2+\varepsilon}k\over {{(2\pi)}^
{2+\varepsilon}}},~~~~~~~\varepsilon = D-4~,
\label{z4}\en
where $D$ is the space-time dimension $(D=4$ for the physical case).
\vskip.3cm
Then we get
\eq
\omega(t) = g^2N{2\over {{(4\pi)}^{D\over 2}}}(-t)^{{D\over 2}-2}{\Gamma
\left(2-{D\over 2}\right)\Gamma^2\left({D\over 2}-1\right)\over{
\Gamma(D-3)}}~.
{}~\label{z5}\en
\vskip.3cm
In the case of pure gluodynamics one loop corrections to the
gluon-gluon-reggeon (GGR), as well as to the reggeon-reggeon-gluon (RRG)
vertices were
calculated by L.N. Lipatov and one of the authors (V.F.)~\cite{FL,FL2}.
\vskip.3cm
In the case of real QCD there is a quark contribution to the vertices; the
quark loop contribution to the GGR vertex was calculated in ref.~\cite{FF}.
Besides that, in the real QCD an extra (compared to the pure gluodynamics
case) vertex appears: the quark-quark-reggeon (QQR) vertex. The existence of
this vertex allows us to check the validity of the assumption that the high
energy behaviour of amplitudes with gluon quantum numbers in the $t$-channel
and negative signature is governed by the Regge pole contribution not only in
the LLA, but beyond it as well. According to this assumption the
amplitude ${\cal A}^{A'B'}_{AB}$ of a process $ A+B\rightarrow
A'+B'$ takes the following factorized form:
\eq
{{\cal P}^-_8}{\cal A}^{A'B'}_{AB} = \Gamma^i_{A'A}{s\over t}\left[({s
\over -t})^{\omega(t)}+({-s\over -t})^{\omega(t)}\right]\Gamma^i_{B'B}~.
\label{z6}\en
Here ${{\cal P}^-_8}$ is the projection operator into the octet colour state
with negative signature, $i$ is the colour index of the reggeized gluon
with the trajectory $j(t)=1+\omega(t)$, given by eq.(\ref{z3}) in the lowest
order of the perturbation theory. For the PPR vertex $\Gamma^i_{A'A}$  in
the helicity basis we get in the lowest order
\eq
\Gamma^{i}_{A'A} = g\langle A'| T^i| A \rangle \delta_{{\lambda_A},{\lambda_
{A'}}}~,
\label{z7}\en
where $\langle A'| T^i| A \rangle$ represents the matrix element of the
group generator in the corresponding representation (i.e. fundamental for
quarks, $T_i=t_i={\lambda_i\over 2}$ and adjoint for gluons, ${(T_i)}_{ab}=
-i{f_{iab}})$ and ${\lambda_A}$ is the helicity of particle $A$.
We assume that the polarization states of the scattered particles are
obtained from those of the initial particles by rotation around the
axis orthogonal to the scattering plane. From eq.(\ref{z6}) we may observe
that the behaviour of the three types of amplitudes (gluon-gluon, quark-quark
and quark-gluon elastic scattering amplitudes) is determined by two vertices,
GGR and QQR; therefore, one of the amplitudes can be expressed in terms of
the others, thus giving a non trivial test of the validity of representation
(\ref{z6}).
\vskip.3cm
Contrary to eq.(\ref{z7}), in higher orders the PPR vertex $\Gamma$ can
contain another spin structure. Due to parity conservation it can be
written in the following form:
\eq
\Gamma^c_{A'A} = g\langle A'| T^i| A \rangle\left[\delta_{{\lambda_A},{
\lambda_{A'}}}(1+\Gamma^{\left(+\right)}_{AA})+\delta_{{\lambda_A},
{-\lambda_{A'}}}\Gamma^{\left(-\right)}_{AA}\right]~,
\label{z8}\en
if relative phases of states with opposite helicity are chosen
appropriately (see ref.s~\cite{FL2,FF} for gluons and eq.(\ref{z36}) below
for quarks). Here $\Gamma^{\left(+\right)}$ and $\Gamma^{\left(-\right)}$
respectively stand for helicity conserving and non conserving loop
contributions to the vertex.
\vskip.3cm
One loop corrections to GGR vertex were calculated in ref.s [3-5].
The contribution of the gluon loop can be written in the form (\ref{z8})
with
\eqn
\Gamma^{\left(+\right)}_{GG}(gluon~loop) = Ng^2{(-t)^{{D\over 2}-2}\over
{(4\pi)^
{D\over 2}}}{\Gamma\left(2-{D\over 2}\right)\Gamma^2\left({D\over 2}-1
\right)\over{\Gamma(D-2)}}\enn
\eq
\times \left\{(D-3)\left[\psi \left(3-{D\over 2}\right)-2\psi \left({D\over
2}-2\right)+\psi (1) \right]-{7\over 4}-{1\over{4(D-1)}}\right\},
\label{z9}\en
\eq
\Gamma^{\left(-\right)}_{GG}(gluon~loop) = Ng^2{(-t)^{{D\over 2}-2}
\over{(4\pi)^
{D\over 2}}}{\Gamma\left(3-{D\over 2}\right)\Gamma^2\left({D\over 2}-1
\right)\over{(D-1)\Gamma(D-2)}},
{}~\label{z10}\en
where $\psi$ is the logarithmic derivative of the gamma function:
\eq
\psi(z) = {{\Gamma'(z)}\over{\Gamma(z)}}~.
\label{z11}\en
For the quark loop contribution we have in turn \cite{FF}
\eq
\Gamma^{\left(^{+}_{-}\right)}_{GG}(quark~loop) = {2g^2\over{(4\pi)^
{D\over2}}}\sum_{f}V^{(f)}_{^{+}_{-}},
{}~\label{z12}\en
where
\eqn
V^{(f)}_{+} = -\Gamma\left(2-{D\over 2}\right)\left[\int^1_0{dxx(1-x)
\over{\left(m_f^2-tx(1-x)\right)^{2-{D\over 2}}}}\right.
\enn
\eqn
\left. + \int^1_0\int^1_0{dx_1dx_2\theta(1-x_1-x_2)
\over{\left(m_f^2-tx_1x_2\right)^{2-{D\over 2}}}}\left({(3-D)\over 2}(2-
x_1-x_2)\right.\right.
\enn
\eq
\left.\left.+(2-{D\over 2}){x_1x_2(1-x_1-x_2)t\over{\left(m_f^2-tx_1x_2
\right)}}\right) \vbox to 27pt{}\right]~,
\label{z13}\en
and
\eq
V^{(f)}_{-} = \Gamma\left(3-{D\over 2}\right)t\int^1_0\int^1_0{dx_1dx_2
\theta(1-x_1-x_2)
x_1x_2(1-x_1-x_2)\over{\left(m_f^2-tx_1x_2\right)^{3-{D\over 2}}}}~.
\label{z14}\en
For massless quarks the two vertices become respectively \cite{FF}
\eq
\Gamma^{\left(+\right)}_{GG}(quark~loop) = {2g^2n_f\over {{(4\pi)}^
{D\over 2}}}(-t)^{{D\over 2}-2}{\Gamma\left(2-{D\over 2}\right)\Gamma^2
\left({D\over 2}\right)\over{\Gamma(D)}},
{}~\label{z15}\en
\eq
\Gamma^{\left(-\right)}_{GG}(quark~loop) = -{2g^2n_f\over {{(4\pi)}^
{D\over 2}}}(-t)^{{D\over 2}-2}{\Gamma\left(3-{D\over 2}\right)\Gamma^2
\left({D\over 2}-1\right)\over{\Gamma(D)}}.
{}~\label{z16}\en
\vskip.3cm
In this paper we calculate the QQR vertex in the one loop approximation and
check the validity of describing amplitudes in terms of the Regge pole
contribution, i.e. representation (\ref{z6}). In section 2 we find the
QQR vertex
by calculating the quark-quark scattering amplitude. Combining the
result obtained
with the one for the GGR vertex, eq.s (\ref{z8})-(\ref{z14}), and Regge
trajectory
(\ref{z5}), we get with the help of eq.(\ref{z6}) a prediction for the
quark-gluon scattering amplitude. In section 3 we perform an independent
calculation of this amplitude and check the consistency of the approach
comparing the result we arrive at with the predicted one. Some conclusions
are illustrated in section 4.

\vskip 0.3cm

{\bf 2. Quark-Quark-Reggeon vertex}

We will extract radiative corrections to the QQR vertex from the amplitude
of quark-quark elastic scattering. The calculation can be carried out
through usual methods
starting from the Feynman diagrams for the quark-quark scattering. However,
for our purposes it
is more convenient to use the method based on the
$t$-channel unitarity relation. This method was used in ref.s~[3-5] to
calculate analogous corrections to the GGR vertex. Here its application
allows us to demonstrate the factorization property of scattering amplitudes
in the most economic way.
\vskip.3cm
{}From the $t$-channel unitarity point of view, it is natural to decompose an
amplitude according to intermediate states in the $t$-channel. In one loop
approximation we need to consider two particle intermediate state in the
$t$-channel. We will schematically represent in Fig. 1 the amplitude of
the elastic
scattering process $ A(p_A)+B(p_B)\rightarrow A'(p_{A'})+B'(p_{B'})$ with
two particle intermediate state in the $t$-channel and will use the
following notations:
\eqn
s = {(p_A+p_B)}^2,~u = {(p_A-p_{B'})}^2,~q = p_A-p_{A'}=p_{B'}-p_B =
p_{C'}-p_C,~t = q^2,
\enn
\eq
s_A = {(p_A+p_C)}^2,~u_A = {(p_A-p_{C'})}^2,~s_B = {(p_B+p_{C'})}^2,~u_B =
{(p_B-p_C)}^2~,
\label{z17}\en
$p_C$ and $p_{C'}$ being the momenta of the two intermediate particles.
\vskip.3cm
Instead of calculating the $t$-channel discontinuities using $2\pi\delta
(p^2-m^2)$ for intermediate particle line, we will calculate the
contribution of diagram in Fig. 1 using full Feynman propagators for
intermediate particles \cite{FL,FF}:
\eq
{\cal A}^{A'B'}_{AB} = \sum_{C,C'}\eta_{CC'}\int{{d^Dp_Cd^Dp_{C'}\delta
^{(D)}(p_C+q-p_{C')}{\cal A}^{A'C'}_{AC}{\cal A}^{B'C}_{BC'}}
\over{({2\pi})^Di\left({p_C}^2-{m_C}^2+i\varepsilon\right)\left({p_{C'}}^2-
{m_C}^2+i\varepsilon\right)}}
{}~.\label{z18}\en
Here, the sum runs over kinds of intermediate particles, their
polarization, colour and flavour states. As already mentioned
in the introduction, the space-time dimension $D$ is not equal to 4, so that
we can use the dimensional regularization for removing both infrared and
ultraviolet divergences~\cite{FL2}. The numerical coefficient $\eta_{CC'}$
depends on the kind of intermediate state in the $t$-channel,
which can be a gluon-gluon or quark-antiquark state: in the first case $\eta_
{GG}={1\over 2}$ because of identity of gluons, while in the second one
$\eta_{Q{\bar Q}}=-1$ because of Fermi statistics. An arbitrary polynomial in
$t$ could be added to the result of integration in eq.(\ref{z18}), because
it does not change the $t$-channel discontinuity, but such
terms would have a wrong asymptotic behaviour incompatible with the
renormalizability of the theory (cf. \cite{FKL,FL2}). Nevertheless, for
massive quarks, some uncertainty still remains. We can add to the r.h.s. of
eq.(\ref{z18}) terms with the pole structure in $t$. In the case of pure
gluodynamics such terms were absent in our regularization scheme because
of lack of appropriate values with mass dimensions . Evidently,
these terms are connected with renormalization and will be
considered at the end of this section.

\vskip 0.3cm

{\bf 2.1 Contribution of the quark-antiquark intermediate state}

Let us first consider the simpler case of the quark-antiquark pair in the $t$-
channel. In this case the amplitudes ${\cal A}^{A'C'}_{AC}$ and ${\cal
A}^{B'C}_{BC'}$ in eq.(\ref{z18}) are the quark-quark scattering amplitudes
taken in Born approximation. For such amplitudes we get
\eqn
{\cal A}^{A'B'}_{AB} = \langle A'| t^c| A \rangle\langle B'| t^c| B \rangle
({g^2\over t})\bar u(p_{A'})\gamma^{\mu}u(p_A)\bar u(p_{B'})\gamma_{\mu}u(
p_B)\enn
\eq
-\delta_{AB}\langle B'| t^c| A \rangle\langle A'| t^c| B \rangle({g^2
\over u})\bar u(p_{B'})\gamma^{\mu}u(p_A)\bar u(p_{A'})\gamma_{\mu}u(
p_B)~,
\label{z19}\en
where the second term contributes only for scattering of identical
particles.
\vskip.3cm
We are interested in the radiative correction to the QQR-vertex when the
reggeon is a reggeized gluon, therefore one needs to project the amplitude
(\ref{z19}) into the octet colour state. Firstly we use the completeness
relation for the generators of the fundamental representation of $SU(N)$:
\eq
t^c_{\alpha\beta}t^c_{\gamma\delta}+{1\over {2N}}\delta_{\alpha\beta}
\delta_{\gamma\delta} = {1\over 2}\delta_{\alpha\delta}\delta_{\beta\gamma}
{}~,
\label{z20}\en
and obtain
\eq
t^c_{\alpha\beta}t^c_{\gamma\delta} = -{1\over N}t^c_{\alpha\delta}t^c_
{\gamma\beta}+{{N^2-1}\over{2N^2}}\delta_{\alpha\delta}\delta_{\beta\gamma}
{}~.\label{z21}\en
Successively, introducing the definition
\eq
{{\cal P}^-_8}{\cal A}^{A'B'}_{AB} = \langle A'| T^c| A \rangle\langle B'|
T^c| B \rangle\left({{\cal A}_8}\right)^{A'B'}_{AB}\label{z22}\en
which will be used in the following, and using the result (\ref{z21})
we have
\eqn
\left({{\cal A}_8}\right)^{A'B'}_{AB} = \bar u(p_{A'})\gamma^{\mu}u(p_A)
({g^2\over t})\bar u(p_{B'})\gamma_{\mu}u(p_B)
\enn
\eq
+{{\delta_{AB}}\over N}
\bar u(p_{B'})\gamma^{\mu}u(p_A)({g^2\over u})\bar u(p_{A'})\gamma_
{\mu}u(p_B)~.
\label{z23}\en
In order to apply the dispersion approach, it is convenient \cite{FL2,FF}
to decompose the amplitudes ${\cal A}^{A'C'}_{AC}$ and ${\cal A}^{B'C}_
{BC'}$ entering eq.(\ref{z18}) into the sum of two
terms which are schematically shown in Fig. 2:
\eq
{\cal A}^{A'C'}_{AC} = {\cal A}^{A'C'}_{AC}(as) + {\cal A}^{A'C'}_
{AC}(na)~,
\label{z24}\en
and analogous expression for ${\cal A}^{B'C}_{BC'}$.
The first term in the r.h.s. of eq.(\ref{z24}) contains the asymptotic
contribution for the Regge kinematics, $s_A\simeq -u_A \gg t$, while the
nonasymptotic part contains the remaining amplitude terms. When
performing the decomposition (\ref{z24}) in the r.h.s. of (\ref{z18}), we
are left with four contributions to ${\cal A}^{A'B'}_{AB}$, corresponding
to the diagrams in Fig. 3. Only the first three of them are important,
instead
the contribution of diagram in Fig. 3$\left(d\right)$ can be disregarded.
In fact, the essential values of variables $s_A$ and $s_B$ are small
$(s_A \sim s_B \sim t)$ for the contribution of this diagram. Consequently,
only transverse (with respect to the $(p_A,p_B)$ plane) components of momenta
$p_C$ and $p_{C'}$ can be taken
in the propagators of intermediate particles, which means that integrals
over $s_A$ and $s_B$  are factorized and can be evaluated by residues;
as a result, the contribution of
diagram in Fig. 3$\left(d\right)$ is purely imaginary in the Regge region and
corresponds to positive signature partial waves \cite{FL2}. Here we are
interested in the radiative corrections to the QQR for the reggeized
gluons, i.e. for the case of negative signature, therefore only diagrams in
Fig.s 3$\left(a,b,c\right)$ can contribute.
\vskip.3cm
The asymptotic contributions take the following form:
\eq
\left({{\cal A}_8}^{(as)}\right)^{A'C'}_{AC} = {(2g^2)\over t}\bar u(p_{A'})
{{\pnot_{B}}\over{s}}u(p_A)\bar u(p_{C'})\pnot_{A}u(p_C)
{}~\label{z25}\en
and
\eq
\left({{\cal A}_8}^{(as)}\right)^{C'B}_{CB'} = {(2g^2)\over t}\bar u(p_{C})
\pnot_{B}u(p_{C'})\bar u(p_{B'}){{\pnot_{A}}\over{s}}u(p_B)~.
\label{z26}\en
We always take a very large value for $s$ (in contrast to $s_A$ and $s_B$,
which are integration variables and can be small as well as large), therefore
we have, in the helicity basis,
\eqn
\bar u(p_{A'}){{\pnot_{B}}\over{s}}u(p_A) = \delta_{{\lambda_A},
{\lambda_{A'}}}~,
\enn
\eq
\bar u(p_{B'}){{\pnot_{A}}\over{s}}u(p_B) = \delta_{{\lambda_B},
{\lambda_{B'}}}~,
\label{z27}\en
where $\lambda_A$ is the helicity of particle $A$. It is assumed that the
polarization states of the scattered particles are obtained from the ones
of the initial particles by rotation around the axis orthogonal to the
scattering plane. That implies
\eq
\varphi'_{\lambda} = exp\left(i\bar{\nu}\cdot\bar{\sigma}{\theta
\over{2}}\right)\varphi_{\lambda}~,~~~~\bar{\nu} = {{{\bar p'}\times
{\bar p}}\over{|{\bar p'}\times\bar p|}}~,
\label{z28}\en
where $\varphi$ and $\varphi'$ are respectively the polarization wave
functions for initial and scattered particles and $\theta$ is the scattering
angle.
\vskip.3cm
Let us note that if we write the asymptotic contribution $\left({{\cal A}_8}^
{(as)}\right)^{A'C'}_{AC}$ (see eq.(\ref{z25})) in the helicity basis for
particles A and A' using the first of eq.s (\ref{z27}), we arrive at the
same expression as for the process in which A and A' are gluons~\cite{FF}.
\vskip.3cm
Taking into account asymptotic parts (\ref{z25}) and (\ref{z26}) and
decomposition (\ref{z24}), from eq.(\ref{z23}) we obtain the following
projections for the nonasymptotic parts into the octet colour state:
\eqn
\left({{\cal A}_8}^{(na)}\right)^{A'C'}_{AC} = g^2\left[\bar u(p_{A'})
\gamma^{\mu}u(p_A){{g^{\mu \nu}_{\perp}}\over t}\bar u(p_{C'})\gamma_
{\nu}u(p_C)\right.\enn
\eq
\left. + {{\delta_{AC}}\over N}\bar u(p_{C'})\gamma^{\mu}u(p_A){1\over u_A}
\bar u(p_{A'})\gamma_{\mu}u(p_C)\right]~,
\label{z29}\en
\eqn
\left({{\cal A}_8}^{(na)}\right)^{CB'}_{C'B} = g^2\left[\bar u(p_{C})
\gamma^{\mu}u(p_C'){{g^{\mu \nu}_{\perp}}\over t}\bar u(p_{B'})\gamma_
{\nu}u(p_B)\right.\enn
\eq
\left. + {{\delta_{CB}}\over N}\bar u(p_{B'})
\gamma^{\mu}u(p_C'){1\over u_B}\bar u(p_{C})\gamma_{\mu}u(p_B)\right]~,
\label{z30}\en
where the decomposition
\eq
g^{\mu \nu} \simeq g^{\mu \nu}_{\perp} + {{2\left(p^{\mu}_Ap^{\nu}_B+
p^{\mu}_Bp^{\nu}_A\right)}\over s}~
\label{z31}\en
has been used and $\perp$ means transverse to the $(p_A,p_B)$ plane. We
neglect terms containing $\bar u(p_{A'})\pnot_{A}u(p_A)$ or $\bar u(p_{B'})
\pnot_{B}u(p_B)$ because they are proportional to the corresponding masses
and therefore cannot give a contribution of order $s$/$t$ to
amplitude (\ref{z18}).
\vskip.3cm
Inserting eq.s (\ref{z25}) and (\ref{z26}) into eq.(\ref{z18}) we get for the
contribution of diagram in Fig. 3$\left(a\right)$ (cf. eq.s (\ref{z20})-
(\ref{z22}) of \cite{FF})
\eq
\left({{\cal A}_8}^{(a)}\right)^{A'B'}_{AB} = \delta_{{\lambda_A},
{\lambda_{A'}}}\delta_{{\lambda_B},{\lambda_{B'}}}\left({2g^2\over t}
\right)^2p^{\mu}_Ap^{\nu}_B\sum_f{\cal P}^{(f)}_{\mu \nu}(q)~,
\label{z32}\en
where summation is performed over quark flavours and
\eqn
{\cal P}^{(f)}_{\mu \nu}(q) = {i\over 2}\int{d^Dp\over{({2\pi})^D}}{
tr\left[\gamma^{\mu}(\pnot+m_f)\gamma^{\nu}(\pnot+\qnot+m_f)\right]
\over{(p^2-m_f^2+i\varepsilon)((p+q)^2-m_f^2+i\varepsilon)}} = \enn
\eq
= {4\Gamma\left(2-{D\over 2}\right)\over {{(4\pi)}^{D\over 2}}}(-g_
{\mu \nu}q^2+q_{\mu}q_{\nu})\int^1_0{{dx}x(1-x)\over {\left(m_f^2-
q^2x(1-x)\right)}^{2-{D\over 2}}}~.
\label{z33}\en
Here, the calculations practically coincide with those of ref.\cite{FF}
because, as we previously noted, the asymptotic contribution $\left({{\cal
A}_8}^{(as)}\right)^{A'C'}_{AC}$ (and $\left({{\cal A}_8}^{(as)}\right)
^{B'C}_{BC'}$ as well) in the helicity basis for particles $A$ and $A'$
(correspondingly $B$ and $B'$) coincides with that of the case considered in
ref.~\cite{FF}, where particles $A$ and $A'$ ($B$ and $B'$) are gluons.
\vskip.3cm
The contribution of diagram in Fig. 3$\left(b\right)$ is expressed by the
product $\left({\cal A}^{(as)}\right)^{A'C'}_{AC}\times \left({\cal A}^{(na)}
\right)^{CB'}_{C'B}$. Using eq.s (\ref{z25}) and (\ref{z30}) and keeping
only terms of order $s$ we find
\eq
\left({{\cal A}_8}^{\left(b\right)}\right)^{A'B'}_{AB} =
\delta_{{\lambda_A},{\lambda_{A'}}}{{g^4s}\over {Nt}}{p^{\mu}_A\over s}
\bar u(p_{B'}){\cal V}_{\mu}(p_B,q)u(p_B)~,
\label{z34}\en
where
\eq
{\cal V}_{\mu}(p_B,q) = i\int{d^Dp\over{({2\pi})^{D}}}{\gamma^\nu
(\pnot+\qnot+m_B)\gamma_\mu(\pnot+m_B)\gamma_{\nu}
\over(p^2-m_B^2+i\varepsilon)((p+q)^2-m_B^2+i\varepsilon)((p_B-p)^2+
i\varepsilon)}~.
\label{z35}\en
The integrals appearing in eq.(\ref{z35}) are calculated in Appendix. Let
us accept that, by definition, states with opposite helicity are connected
by relation
\eq
\varphi_{-\lambda} = \bar{\nu}\cdot\bar{\sigma}\varphi_{\lambda}~,
{}~~~~\bar{\nu} = {{{\bar p'}\times{\bar p}}\over{|{\bar p'}\times
\bar p|}}~,
\label{z36}\en
which, in the helicity basis, leads to
\eq
\bar u(p_{A'})u(p_A) = 2m_A\delta_{{\lambda_A},{\lambda_{A'}}}-
i\sqrt{-t}\delta_{{\lambda_A},{-\lambda_{A'}}}~.
\label{z37}\en
With such a definition we obtain
\eqn
{p^{\mu}_A\over s}\bar u(p_{B'}){\cal V}_{\mu}(p_B,q)u(p_B) =
\enn
\eqn
{{\Gamma\left(3-{D\over 2}\right)}\over{({4\pi})^{D\over 2}}}\int^1_0
{dx\over{\left(m_B^2-x(1-x)t\right)^{3-{D\over 2}}}}\left\{{2\delta_
{{\lambda_B},{\lambda_{B'}}}\over{4-D}}\left[t\left({1\over{D-3}}\right.
\right.\right.
\enn
\eq
\left.\left.\left. +(D-3)x(1-x)\right)-{D-1\over{D-3}}m^2_B\right]
-i\delta_{{\lambda_B},{-\lambda_{B'}}}{5-D\over{D-3}}m_B\sqrt{-t}
\vbox to 17,5pt{}\right\}~.
\label{z38}\en
Let us pay attention to the fact that the contribution given by eq.s
(\ref{z34}) and (\ref{z36}) does not depend on the nature of particle A.
\vskip.3cm
Finally the contribution $\left({{\cal A}_8}^{\left(c\right)}\right)^
{A'B'}_{AB}$ of the diagram in Fig. 3$\left(c\right)$ can be obtained
from eq.s (\ref{z34}) and (\ref{z38}) by the substitution
$A \leftrightarrow B$.
Summing up the three contributions and comparing the sum ${\cal A}^{\left(a
\right)}+{\cal A}^{\left(b\right)}+{\cal A}^{\left(c\right)}$ with
the factorized form (\ref{z6}), where the vertices are
defined through eq.(\ref{z8}), we find
\eqn
\Gamma^{\left(+\right)}_{QQ}(q\bar q~state) = {g^2\over{(4\pi)^{D\over2}}}
\Gamma\left(2-{D\over 2}\right)\left\{ \vbox to 26.66pt{}-2\sum_{f}
\int^1_0{dxx(1-x)\over{(m_f^2
-x(1-x)t)^{2-{D\over 2}}}}\right.
\enn
\eq
\left.+ {1\over 2N}\int_0^1{dx\over{\left(m^2_Q-x(1-x)t\right)}^
{3-{D\over 2}}}
\left[t\left({1\over D-3}+(D-3)x(1-x)\right)-m_Q^2{(D-1)\over (D-3)}
\right]\right\}~,
\label{z39}\en
\eq
\Gamma^{\left(-\right)}_{QQ}(q\bar q~state) = {-ig^2\over {(4\pi)}^
{D\over 2}}{\Gamma\left(3-{D\over 2}\right) \over 2N}m_Q\sqrt{-t}
{(5-D)\over (D-3)}\int_0^1{dx\over {(m_Q^2-x(1-x)t)}^{3-{D\over 2}}}~.
\label{z40}\en
We expect that for massless quarks only the helicity conserving part of the
vertex survives and in fact in this case eq.s (\ref{z39}) and (\ref{z40})
reduce to
\eqn
\Gamma^{\left(+\right)}_{QQ}(q\bar q~state)|_{m_q=0} = {g^2\over{(4\pi)^
{D\over2}}}(-t)^{{D\over 2}-2}\left[\vbox to 23.3pt{}-2\Gamma\left(2-
{D\over 2}\right)n_f{\Gamma^2\left({D\over 2}\right)\over{\Gamma(D)}}
\right.
\enn
\eq
\left. +{\Gamma\left(3-{D\over 2}\right)\over{2N}}{\Gamma^2\left({D\over 2}-1
\right)\over{\Gamma(D-2)}}\left(2+{D\over{2\left({D\over 2}-2\right)^2}}
\right)\right]~,
\label{z41}\en
\eq
\Gamma^{\left(-\right)}_{QQ}(q\bar q~state)|_{m_q=0} = 0~.
\label{z42}\en
Here $n_f$ is the number of quark flavours.

\vskip 0.3cm

{\bf 2.2. Contribution of the two gluon intermediate state}

Now let us consider the contribution of two gluon intermediate state in the
$t$-channel to the QQR vertex. The general lines of the consideration are
the same as before, but now we need to take the quark-gluon scattering
amplitude in Born approximation for the amplitudes ${\cal A}^{A'C'}_{AC}$
and ${\cal A}^{CB'}_{C'B}$ in eq.(\ref{z18}). Again, as in the quark
contribution case, we are interested only in parts of the amplitudes which
correspond to the octet colour state in the $t$-channel, because our
reggeon is the reggeized gluon. Moreover, we need to keep only the F-type
colour octet for intermediate gluons, because only this colour state
survives in the Regge asymptotic regime. Consequently the asymptotic
contribution
${\cal A}^{(as)}$ in the decomposition (\ref{z24}) can contain only
this colour state, which is,
therefore, the only state that contributes to the essential diagrams in
Fig.s 3$\left(a,b,c\right)$.
\vskip.3cm
This part of the quark-gluon scattering amplitude ${\cal A}^{A'C'}_{AC}$ can
be written as (cf. ref.\cite{FF})
\eqn
\left({{\cal A}_8}\right)^{A'C'}_{AC} = {(-g^2)\over 2}\bar u(p_{A'})
\left\{\enot_C{\pnot_A - \pnot_{C'} + m_A\over{u^2_A-m^2_A}}
\enot^*_{C'} -
\enot^*_{C'}{\pnot_A + \pnot_{C} + m_A\over{s^2_A-m^2_A}}
\enot_C\right.
\enn
\eq
\left.+ {2\over t}\left[(\pnot_C+\pnot_{C'})e^*_{C'}\cdot e_C +2\enot_C
(e^*_{C'}\cdot q) - 2\enot^*_{C'}(e_C\cdot q\right] \vbox to 16.66pt{}
\right\}u(p_A)~.
\label{z43}\en
In contrast with ref. \cite{FF}, here gluons are intermediate particles, so
that we do not fix their gauge. Instead we write the asymptotic parts of
the amplitudes in a form similar to the one used in ref. \cite{FL2} for
the gluon-gluon scattering amplitude. The possibility of such a choice comes
from the fact that the high energy behaviour of both amplitudes is
determined by the $t$-channel exchange of a gluon which reggeizes. In the
helicity basis (see eq.(\ref{z27})) we have
\eq
\left({{\cal A}_8}^{(as)}\right)^{A'C'}_{AC} = g^2\delta_{{\lambda_A},
{\lambda_{A'}}}e^{*{\sigma'}}_{C'}e^{\sigma}_C\Gamma_
{\sigma\sigma'}\left(p_C,p_{C'},p_A\right)~,
\label{z44}\en
\eq
\left({{\cal A}_8}^{(as)}\right)^{CB'}_{C'B} = g^2\delta_{{\lambda_B},
{\lambda_{B'}}}e^{*{\sigma}}_Ce^{\sigma'}_{C'}\Gamma_{\sigma'\sigma}
\left(p_{C'},p_C,p_B\right)~,
\label{z45}\en
with
\eqn
\Gamma^{\sigma\sigma'}\left(p_C,p_{C'},p_A\right) = -g^{\sigma\sigma'}
{\left(s_A-u_A\right)\over t} - 2\left({2\over t}+{1\over{s_A-m^2_A}}
\right)p^{\sigma}_Aq^{\sigma'}
\enn
\eq
+2\left({2\over t}+{1\over{u_A-m^2_A}}\right)p^{\sigma'}_Aq^{\sigma}+
2\left({1\over{s_A-m^2_A}}-{1\over{u_A-m^2_A}}\right)p^{\sigma}_Ap^
{\sigma'}_A~
\label{z46}\en
and
\eq
q = p_{C'}-p_{C}~,~~~~s_A = {\left(p_A+p_C\right)}^2~,~~~~u_A =
{\left(p_A-p_{C'}\right)}^2~.
\label{z47}\en
It is worth to note that the tensor $\Gamma^{\sigma\sigma'}$ can
be considered as the
generalization of the corresponding tensor used in ref. \cite{FL2}, for the
gluon-gluon scattering amplitude, to the case for which $p^2_A=p^2_{A'}=
m^2_A\neq0$.
\vskip.3cm
In correspondence to the form we choose for the asymptotic term, the
nonasymptotic contribution ${\cal A}^{(na)}$, taking into account
eq.(\ref{z27}), becomes
\eq
\left({{\cal A}_8}^{(na)}\right)^{A'C'}_{AC} = {(-g^2)\over 2}e^
{*\sigma'}_{C'}e^{\sigma}_C\bar u(p_{A'}){\chi}_{\sigma\sigma'}
(p_C,p_{C'};p_A,p_B)u(p_A)~,
\label{z48}\en
\eq
\left({{\cal A}_8}^{(na)}\right)^{CB'}_{C'B} = {(-g^2)\over 2}e^
{*\sigma}_Ce^{\sigma'}_{C'}\bar u(p_{B'}){\chi}_{\sigma'\sigma}
(p_{C'},p_C;p_B,p_A)u(p_B)~,
\label{z49}\en
where
\eqn
{\chi}^{\sigma\sigma'}(p_C,p_{C'};p_A,p_B) =
\enn
\eqn
\left(\gamma^{\varrho}-{2\pnot_Bp^{\varrho}_A\over s}\right)\left[2g^
{\sigma\sigma'}{(p_C+p_{C'})^{\varrho}\over t}+2g^{\sigma\varrho}\left
({2q^{\sigma'}\over t}-{{(p_A-q)}^{\sigma'}\over{s_A-m^2_A}}\right)\right.
\enn
\eq
\left.-2g^{\sigma'\varrho}\left({2q^{\sigma}\over t}-{{(p_A-q)}^{\sigma}
\over{u_A-m^2_A}}\right)\vbox to 20pt{}\right]- {\gamma^{\sigma}\pnot_C
\gamma^{\sigma'}
\over{u_A-m^2_A}}-{\gamma^{\sigma'}\pnot_{C'}\gamma^{\sigma}\over{s_A-
m^2_A}}~.
\label{z50}\en
Let us pay attention that the tensor $\Gamma_{\sigma\sigma'}\left(p_C,p_{C'},
p_A\right)$ has the following properties on the mass-shell $p^2_C=p^2_
{C'}=0$:
\eq
\Gamma_{\sigma\sigma'}\left(p_C,p_{C'},p_A\right)p^{\sigma}_C = \Gamma_
{\sigma\sigma'}\left(p_C,p_{C'},p_A\right)p^{\sigma'}_{C'} = 0~.
\label{z51}\en
Consequently we can use the Feynman summation over polarization states of
intermediate gluons:
\eq
\sum_{\lambda}e^{(\lambda)}_{\mu}e^{*(\lambda)}_{\nu} =
-g_{\mu\nu}~,
\label{z52}\en
when we calculate the contributions of the diagrams in Fig.s
3$\left(a,b,c\right)$, without
introducing the Faddeev-Popov ghosts.
\vskip.3cm
In the case of the diagram in Fig. 3$\left(a\right)$ we need to calculate
the product of
the asymptotic parts (\ref{z44}) and (\ref{z45}). The essential region of
integration in eq.(\ref{z18}) for large $s$ and fixed $t$ in this case is
determined by relations
\eq
p^2_C \sim p^2_{C'}\sim {s_As_B\over s}\sim t~;~~~~t\la s_A\sim u_A
\la s~;~~~~t\la s_B\sim u_B\la s~.
\label{z53}\en
In this region from eq.(\ref{z46}) we find
\eqn
\Gamma^{\sigma\sigma'}\left(p_C,p_{C'},p_A\right)\Gamma_{\sigma'\sigma}
\left(p_{C'},p_C,p_B\right) = {(D-4)\over{t^2}}\left(u_A-s_A\right)\left
(u_B-s_B\right)
\enn
\eqn
+{4\over{t^2}}\left(s_As_B+u_Au_B\right)+{16s\over t}+s^2\left({1\over
{s_A-m^2_A}}-{1\over{u_A-m^2_A}}\right)\left({1\over{s_B-m^2_B}}-{1\over
{u_B-m^2_B}}\right)
\enn
\eqn
+4s\left({1\over{s_A-m^2_A}}+{1\over{s_B-m^2_B}}+{1\over{u_A-m^2_A}}+{1\over
{u_B-m^2_B}}\right)-2\left({{u_B-m^2_B}\over {s_A-m^2_A}}+{{s_B-m^2_B}\over
{u_A-m^2_A}}\right)
\enn
\eq
\times\left(1-{2m^2_A\over t}\right)-2\left({{u_A-m^2_A}\over{s_B-
m^2_B}}+{{s_A-m^2_A}\over{u_B-m^2_B}}\right)\left(1-{2m^2_B\over t}\right)~.
\label{z54}\en
As to be expected, this expression differs from the corresponding one of ref.
\cite{FL2} only by mass terms, as well as $\Gamma^{\alpha\beta}$ does.
\vskip.3cm
The integrals appearing in eq.(\ref{z18}), after substitution of
eq.s (\ref{z44}), (\ref{z45}) and (\ref{z54}), are presented in Appendix.
By means of these integrals one
can give the contribution of the diagram in Fig. 3$\left(a\right)$
 the following form:
\eq
\left({{\cal A}_8}^{\left(a\right)}\right)^{A'B'}_{AB} =
{Ng^4\over{(4\pi)^{D\over2}}}
\delta_{{\lambda_A},{\lambda_{A'}}}\delta_{{\lambda_B},{\lambda_{B'}}}
{2s\over t}\left[a(s,t)+\Delta a(m^2_A,t)+\Delta a(m^2_B,t)\right]~.
\label{z55}\en
Here $a(t)$ is the contribution for the massless case (it is the same
as in ref.\cite{FL2}):
\eqn
a(t) = { {\Gamma\left(2-{D\over 2}\right)\Gamma^2\left({D\over 2}-1\right)}
\over{(-t)^{2-{D\over 2}}\Gamma(D-2)} }\left\{\vbox to 16.66pt{}(D-3)
\left[ln\left({-s\over -
t}\right)+ln\left({-u\over -t}\right)\right.\right.
\enn
\eq
\left.\left. + 2\psi \left(3-{D\over 2}\right)-4\psi \left({D\over 2}-2\right)
+2\psi (1) \right]+{1\over{2(D-1)}}-{4\over{D-4}}-{9\over 2}\right\}~,
\label{z56}\en
where $\psi (z)$ is defined in eq.(\ref{z11}), while  ${\Delta}
a(m^2,t)$ is mass dependent and becomes zero at $m=0$:

\eqn
{\Delta}a(m^2,t) = \Gamma\left(3-{D\over 2}\right)\int^1_0\int^1_0
dx_1dx_2\theta(1-x_1-x_2)
\enn
\eqn
\left[{t\left(1-x_1\right)^2\over x_1}\left({1\over \left(x^2_1m^2-x_2
(1-x_1-x_2)t\right)^{3-{D\over 2}}}-{1\over \left((-t)
\left(1-x_1-x_2\right)\right)^{3-{D\over 2}}}\right)\right.
\enn
\eq
\left.-{2m^2x_1\over \left(x_1^2m^2-x_2\left(1-x_1-x_2t\right)\right)^
{3-{D\over 2}}}\right]~.
\label{z57}\en
The essential point in eq.(\ref{z50}) is the separation of dependences on
$m_A$ and $m_B$. Such separation is a vital necessity for the interpretation
of the amplitude in terms of the Regge pole contribution (see eq.(\ref{z6})).
\vskip.3cm
Let us evaluate the contribution of the diagram in Fig. 3$\left(b\right)$.
It is
expressed in terms of the integral (\ref{z18}), where we need to use the
amplitudes (\ref{z44}) and (\ref{z49}).
The essential region of integration in eq.(\ref{z18}) for this case is
\eq
p_C^2\sim p^2_{C'}\sim s_B\sim u_B\sim t,~~~~s_A\sim u_A\sim s.~
\label{z58}\en
In this region from (\ref{z46}) and (\ref{z50}) we obtain
\eqn
\Gamma^{\sigma\sigma'}(p_C,p_C',p_A)\bar u(p_B')
\chi_{\sigma'\sigma}(p_{C'},p_C;p_B,p_A)u(p_B) =
\enn
\eqn
\bar u(p_B')\left\{{s_A-u_A\over t}\left[{2\over
s_B-m_B^2}\left(m_B-\left(2m_B^2-t\right){\pnot_A\over s}\right)
\right.\right.
\enn
\eqn
\left.\left.-{2\over u_B-m_B^2}\left(m_B-\left(2m_B^2-t\right)
{\pnot_A\over s}\right)-(D-2)\left({\pnot_{C'}\over u_B-m_B^2}+
{\pnot_C\over s_B-m_B^2}\right)\right]\right.
\enn
\eq
\left.+{4\over t}\left[{\qnot \pnot_{C'}\pnot_A-\pnot_A\pnot_{C'}
\qnot\over u_B-m_B^2}+{\pnot_A\pnot_C\qnot-\qnot\pnot_C\pnot_A\over
s_B-m_B^2}\right]\right\}~.
\label{z59}\en
An important property of this expression is its independence on $m_A$.
Remembering that the tensors $\Gamma^{\sigma\sigma'}\left(p_C,p_{C'},p_A
\right)$ for the cases when the particles $A$ and $A'$ are gluons or quarks
differ only by mass terms, we conclude that the contribution of diagram in
Fig. 3$\left(b\right)$ does not depend on the nature of particles $A, A'$
if it is written
in the helicity state basis for these particles. With the help of the
Appendix, where integrals appearing in eq.(\ref{z18}) after substituting
eq.s (\ref{z44}), (\ref{z49}) and (\ref{z59}) are  calculated, and using
the relation
\eq
\bar u\left(p_{B'}\right)\left(\qnot\pnot_B\pnot_A-\pnot_A\pnot_B
\qnot\right)u\left(p_B\right) = \bar u\left(p_{B'}\right)\left[\left(
t-4m_B^2\right)\pnot_A+2sm_B\right]u\left(p_B\right)~,
\label{z60}\en
which is a result of a simple algebra, we obtain for this contribution
\eqn
\left(A_8^{\left(b\right)}\right)^{BB'}_{AA'} =
\delta_{{\lambda_A},{\lambda_{A'}}}{2s\over t}
{g^4N\Gamma
\left(3-{D\over2}\right)\over (4\pi)^{D\over 2}}\bar u
(p_{B'})\int_0^1\int_0^1{dx_1dx_2\theta\left(1-x_1-x_2\right)\over{\left
(x_1^2m_B^2-x_2(1-x_1-x_2)t\right)^{3-{D\over 2}}}}
\enn
\eqn
\left[{\pnot_A\over s}\left(-2m_B^2x_1-{D-2\over D-4}\left(m_B^2x_1^2
-x_2\left(1-x_1-x_2\right)t\right)\right)\right.
\enn
\eq
\left. +m_B\left(x_1-\left({D\over 2}-1\right)x_1^2\right)
\vbox to 16,66pt{}\right]u(p_B)~.
\label {z61}\en
\vskip .3cm
The contribution of the diagram in Fig. 3$\left(c\right)$ can be obtained
from eq.(\ref{z61}) by the substitution $A\leftrightarrow B$,
$A'\leftrightarrow B'$.
The total contribution of the two gluon intermediate state in the $t$-channel
to the asymptotic of the quark-quark scattering amplitude with the octet color
state and negative signature in the $t$-channel is given by the sum
${\cal A}^{\left(a\right)}+{\cal A}^{\left(b\right)}+{\cal A}
^{\left(c\right)}$. Comparing it with
representation (\ref{z6}) we are once more convinced that Regge trajectory
$\omega$ in the lowest order is given by eq.(\ref{z5}) and find the two gluon
intermediate state contribution to the quark-quark-reggeon vertex. For the
helicity conserving part of this contribution (see eq.(\ref{z8})),
performing the decomposition
\eq
\Gamma^{\left(+\right)}_{QQ} = \Gamma^{\left(+\right)}_{QQ}\mid_{m_Q=0}+
\Delta\Gamma^{\left(+\right)}_{QQ}~,
\label{z62}\en
with the help of eq.s (\ref{z27}) and (\ref{z37}) we obtain
\eqn
{\Gamma^{\left(+\right)}_{QQ}\left(gg~state\right)}\mid_{m_Q=0} = {g^2N
\over \left(4\pi\right)^{D\over 2}}{\Gamma\left(2-{D\over 2}\right)
\Gamma^2\left({D\over 2}-1\right)\over \left(-t\right)^{2-{D\over 2}}
\Gamma\left(D-2\right)}\left[\vbox to 16.66pt{}\left(D-3\right)
\left(\psi \left(3-{D\over 2}\right)\right.\right.
\enn
\eq
\left.\left. -2\psi \left(
{D\over 2}-2\right)+\psi \left(1\right)\right)+{1\over 4\left(D-1\right)}
-{2\over D-4}-{7\over 4}\right]~,
\label{z63}\en
and
\eqn
\Delta\Gamma_{QQ}^{\left(+\right)}\left(gg~state\right) = {g^2N
\over \left(4\pi\right)^{D\over 2}}\Gamma\left(3-{D\over 2}\right)
\int_0^1\int_0^1dx_1dx_2\theta\left(1-x_1-x_2\right)
\enn
\eqn
\left[t\left({\left(1-x_1\right)^2\over x_1}+{\left(D-2\right)\over
\left(D-4\right)}x_2\left(1-x_1-x_2\right)\right)\left({1\over
\left(x_1^2m^2_Q-x_2\left(1-x_1-x_2\right)t\right)^{3-{D\over 2}}}
\right.\right.
\enn
\eq
\left.\left. -{1\over\left(-x_2\left(1-x_1-x_2\right)t\right)^
{3-{D\over 2}}}\vbox to 27.5pt{}\right)-{m_Q^2\left(2x_1+(D-2)(D-3)
{x_1^2\over D-4}\right) \over \left(x_1^2m_Q^2-x_2
\left(1-x_1-x_2\right)t\right)^{3-{D\over 2}}}\right]~.
\label{z64}\en
\vskip 0.3cm
For the helicity non-conserving part we have
\eqn
\Gamma^{\left(-\right)}_{QQ}\left(gg~state\right) = -i{g^2N
\over \left(4\pi\right)^{D\over 2}}\Gamma\left(3-{D\over 2}\right)
m_Q\sqrt{-t}
\enn
\eq
\times\int_0^1\int_0^1{dx_1dx_2\theta\left(1-x_1-x_2\right)\over
\left(x_1^2m_Q^2-x_2\left(1-x_1-x_2\right)t\right)^{3-{D\over 2}}}
\left(x_1-\left({D\over 2}-1\right)x_1^2\right)~.
\label{z65}\en
It vanishes in the zero mass limit, as it should do.

\vskip 0.3cm

{\bf 2.3. Renormalization of the QQR vertex}

It was explained at the beginning of this section that we can add a term
with the pole structure in $t$ to the r.h.s. of eq.(\ref{z18}) without
changing the $t$-channel discontinuity and without spoiling the
renormalizability of the theory. That means we can add an expression which
is equal to the Born amplitude with some constant coefficient. In
principle, we could put on the helicity conserving part of the QQR vertex
some condition (of the type of $\Gamma^{\left(+\right)}_{QQ}\mid_{t=
-{\mu}^2}=0$
which would be a definition of the renormalized coupling constant. But it is
useful to have a possibility to express the $QQR$ vertex in terms
of the coupling constant in commonly used renormalization schemes, such as
the $\overline{MS}$ scheme. In order to have such a possibility we need
to connect
our results to those which are obtained by the usual approach, in terms of
Feynman diagrams. In the diagrammatic approach the terms under discussion
may come from quark-gluon vertices and gluon polarization operator at
$t=0$ and from self-energy insertions into external quark legs. It is easy
to observe that the first two contributions in Feynman gauge
\footnote{One can wonder why Feynman gauge did appear
here, while we
used gauge invariant amplitudes and were talking about calculations without
introducing the Faddev-Popov ghosts. The matter is, the gauge invariance
used (see for example eq.(\ref{z51})) is held on the mass shell only and has
no relation to the terms under consideration.}are taken
into account properly in eq.s (\ref{z32}), (\ref{z34}) and
(\ref{z55}), (\ref{z61}). So, we need only to consider
the self-energy insertions into external quark legs. It means,
that in the one loop approximation we need to add to the helicity
conserving part of the $QQR$ vertex the value
\eq
\Gamma_{QQ}^{\left(+\right)}(self-energy) = -{\partial\Sigma(\pnot)\over
\partial \pnot}\mid_{\pnot=m_Q}~,
\label{z66}\en
where $\Sigma(\pnot)$ is the mass operator of the quark:
\eq
\Sigma(\pnot) = -g^2{\left(N^2-1\right)\over{2N}}\int{d^Dk\over({2\pi})^Di}
{\gamma^{\mu}\left(\pnot-\knot+m_Q\right)\gamma_{\mu}\over{\left(k^2+i
\varepsilon\right)\left((p-k)^2-m^2_Q+i\varepsilon\right)}}~.
\label{z67}\en
An elementary calculation gives
\eq
\Gamma_{QQ}^{\left(+\right)}(self-energy) = g^2{\left(N^2-1\right)\over{2N}}
{\Gamma\left(2-{D\over 2}\right)\over{{(4\pi)}^{D\over 2}}}m^{D-4}_Q
\left({1-D\over{D-3}}\right)~.
\label{z68}\en
\vskip.3cm
The total one loop correction to the helicity conserving part of the QQR
vertex is given by the sum
\eq
\Gamma_{QQ}^{\left(+\right)} = \Gamma_{QQ}^{\left(+\right)}(q\bar q~state)
+\Gamma_{QQ}^{\left(+\right)}(gg~state)+\Gamma_{QQ}^{\left(+\right)}
(self-energy)~,
\label{z69}\en
where the first term in the r.h.s. is given by eq.(\ref{z39}), the second
by eq.(\ref{z62}) and the third by eq.(\ref{z68}). For the helicity non
conserving part we have only contributions of the first two types, which are
given by eq.s (\ref{z40}) and (\ref{z65}) correspondingly.
\vskip.3cm
In all the above formulas we used the nonrenormalized coupling constant $g$.
Therefore expression (\ref{z69}) for $D\rightarrow 4$ contains
singularities coming from ultraviolet as well as infrared divergences of
Feynman integrals. Let us note here that the helicity non conserving
part of the vertex does not contain the ultraviolet divergences. We can
remove the ultraviolet divergences
expressing $g$ in terms of the renormalized coupling constant, for example,
in the $\overline{MS}$ scheme:
\eq
g = g_{\mu}{\mu}^{4-D\over 2}\left\{1+\left({11\over 3}N-{2\over 3}n_f
\right){g^2_{\mu}\over{{(4\pi)}^{D\over 2}(D-4)}}\left[1+{D-4\over 2}
\left(\ln{1\over{4\pi}}-\psi(1)\right)\right]+\cdot\cdot\cdot\right\}~,
\label{z70}\en
where $g_{\mu}$ is the renormalized coupling constant at the normalization
point $\mu$ and $\psi(z)$ has been already defined (see eq.(\ref{z11})).

\vskip 0.3cm

{\bf 3. Check of the approach consistency}

Now we have both the gluon-gluon-reggeon vertex $\Gamma^R_{GG}$ and
quark-quark-reggeon vertex $\Gamma_{QQ}^R$ in the one loop approximation.
The first of them was calculated by using the gluon-gluon scattering amplitude
as a tool, and the second via the quark-quark scattering amplitude. The
process of quark-gluon scattering has not been considered up to now. But the
reggeon contribution to this amplitude (see eq.(\ref{z6})) is expressed
in terms of the $GGR$ vertices and $QQR$ vertices and the gluon trajectory
$1+\omega\left(t\right)$, which is given in the one loop approximation by
the formula (\ref{z5}). It allows us to check the validity of
representation (\ref{z6}) for
the high energy behavior of the amplitudes with gluon quantum numbers in
the $t$-channel and negative signature.\par
It can be done by calculating the quark-gluon scattering amplitude and
comparing the results of calculation with the expression given by formula
(\ref{z6}) in terms of known trajectory  $1+\omega\left(t\right)$ and vertices
$\Gamma_{QQ}^R$ and $\Gamma_{GG}^R$. But really we need not to perform any
new calculation. The method of calculations used for gluon-gluon and
quark-quark scattering amplitudes allows us to check the validity of the
expression (\ref{z6}) simply by keeping an eye on the calculations we
performed. Our starting point in calculating any amplitude is
eq.(\ref{z18}). In each case
we have two gluon and quark-antiquark intermediate states in the
$t$-channel. An essential step in the calculation is the decomposition of the
amplitude entering in the integrand of eq.(\ref{z18}) into the sum
(\ref{z24}) (see Fig. 2) of asymptotic and non asymptotic parts. An
important fact is that a product of non asymptotic parts in the integrand
in eq.(\ref{z18}) cannot give
a contribution of order $s$ to an amplitude with gluon quantum numbers in
the $t$-channel and negative signature which we are interested in, so we are
left with contributions presented schematically in Fig. 3$\left(a,b,c
\right)$.
\vskip.3cm
The next important fact is that the asymptotic parts of the amplitudes
$\left(A^{\left(as\right)}_8\right)_{AC}^{A'C'}$ and $\left(A_8^
{\left(as\right)}\right)_{BC'}^{B'C'}$ can be chosen in a factorized form.
 Of course, this fact is strictly related to the case in which the high
energy behaviour of the amplitudes is determined by gluon exchange in the
$t$-channel. We choose the asymptotic parts for the case of quark-antiquark
intermediate state $\left(C C'\right)$ in the $t$-channel in the following
form:
\eq
\left(A_8^{\left(as\right)}\right)_{AC}^{A'C'} = {2g^2\over t}\delta_
{\lambda_A\lambda_{A'}}\bar u\left(p_{C'}\right)\pnot_Au\left(p_C\right)~
\label{z71}\en
in the helicity basis for particles $A$ and $A'$, independently of what
they are:
gluons (see eq.s (\ref{z17}), (\ref{z13}) of Ref.\cite{FF}) or quarks (see
eq.s (\ref{z25}) and (\ref{z27}) of this
paper). Therefore, in this intermediate channel the contribution of the
diagram in Fig. 3$\left(a\right)$ does not depend on the kind of particles
$A,A'$ and
$B,B'$ (see eq.(\ref{z32})). For the same reason the contribution of the
diagram in Fig. 3$\left(b\right)$ (Fig. 3$\left(c\right)$) depends only on the
kind of the particles $B,B'$
($A,A'$). Taking into account that these diagrams contribute only to the
vertices $\Gamma^R_{BB'}$ and $\Gamma^R_{AA'}$ correspondingly, we conclude
that the contribution of the quark-antiquark intermediate state in the
$t$-channel to an amplitude of any of the processes under consideration
can be put in the form of eq.(\ref{z6}), where the vertices
$\Gamma^R_{AA'}$
($\Gamma^R_{BB'}$) do not depend on the kind of particles $B,B'$
($A, A'$).
\vskip 0.3cm
For the case of the two gluon intermediate state the conclusion is the
same, although the properties of the asymptotic contributions are slightly
changed. We choose these contributions in the form of
eq.s (\ref{z44})-(\ref{z46}) and
here the dependence on the kind of the particles $A, A'$ ($B, B'$) enters
through the masses of these particles\footnote{Let us note, that, on the
contrary, the
dependence on $p_A^2=m_A^2$ is negligible in eq.(\ref{z66})}. Of course
in the Regge asymptotic limit for the amplitude
$A_{AC}^{AC'}$ ($A_{BC'}^{B'C})$, that
means for $s_A\sim u_A\gg t$ ($s_B\sim u_B\gg t$), this dependence becomes
negligible, but we need to integrate over $s_A$ ($u_A$) in eq. (18).
Therefore we choose these asymptotic contributions in such a form, which
conserve the analytic properties of the exact amplitudes.
\vskip.3cm
An essential property of asymptotic parts (\ref{z44}) and (\ref{z45}))
 is that the
contribution of the diagram in Fig. 3$\left(a\right)$, being calculated in
terms of these parts,
is presented in the form of eq.(\ref{z55}), where the dependences on
the masses
$m_A$ and $m_B$ are separated. Therefore, all the dependence on the
kind of particles $A,A'$ ($B, B'$) coming from this contribution is
included into the vertices $\Gamma_{AA'}^R$ ($\Gamma_{BB'}^R$). The same is
true for the case of the contributions of the diagrams in Fig. 3$\left(b,c
\right)$ as well,
because the contribution of the diagram in Fig. 3$\left(b\right)$
(Fig. 3$\left(c\right)$) depends only on
the kind of particles $B, B'$ ($A,A'$) (see eq.(\ref{z61})),
just as in the case
of the quark-antiquark intermediate state. Consequently we come to the same
conclusion as for this case. Since the contributions of the quark-antiquark
and two gluons intermediate states enter into the $PPR$-vertices
additively, it means that the high energy behaviour of all QCD elastic
scattering
amplitudes with gluon quantum numbers in the $t$-channel and negative
signature are presented by the Regge pole contribution (6).

\vskip 0.3cm

{\bf 4. Conclusions}

We calculated one loop corrections to the quark-quark-reggeon vertex in the
QCD, where the reggeon is a reggeized gluon. Taking into account this vertex
together
with the gluon-gluon-reggeon vertex calculated before, we get Regge pole
contributions to gluon and quark elastic scattering processes. Since
non-logarithmic terms of these contributions to the amplitudes of the three
processes
(gluon-gluon, quark-quark and quark-gluon elastic scattering) are
expressed in terms of the two vertices, these amplitudes have to satisfy
non trivial relations if the Regge pole only contributes to large $s$
behaviour of the amplitudes with gluon quantum numbers and negative
signature in the $t$-channel. We have checked that these relations are
fulfilled, i.e. the representation (\ref{z6}) of these amplitudes in terms
of the Regge pole contribution is applicable beyond the leading logarithmic
approximation (LLA) for the Regge region.
The results obtained are needed for the next step in the calculation
program \cite{LF} for corrections to the LLA: calculation of two loop
corrections to the gluon Regge trajectory. In the two loop approximation a
part of corrections to the trajectory comes from two particle intermediate
states in the two channels. The calculation of this part can be performed by
using the presented results. It will be done in a subsequent
publication.

\vskip 2.5cm

{\bf Appendix}
\vskip.5cm
\renewcommand{\theequation}{A.\arabic{equation}}
\setcounter{equation}{0}

For the reader's convenience, here we are presenting the Feynman integrals
appearing in eq.(\ref{z18}).
\vskip.3cm
Let us first consider the case of quark-antiquark intermediate state.
Calculating the contribution of the diagram in Fig.3$\left(a\right)$
we meet the followings integrals:
\eqn
I_{Q2} = -i\int{d^Dp\over{\left(p^2-m^2+i\varepsilon\right)\left((p+q)^2-
m^2+i\varepsilon\right)}} =
\enn
\eq
{\pi}^{D\over 2}\Gamma\left(2-{D\over 2}\right)\int_0^1{dx\over{\left(m^2-x
(1-x)t\right)^{2-{D\over 2}}}}~,
\en
\eq
I^{\mu}_{Q2} = -i\int{d^Dp\left(-p^{\mu}\right)\over{\left(p^2-m^2+i
\varepsilon\right)\left((p+q)^2-m^2+i\varepsilon\right)}} = {q^{\mu}
\over 2}I_{Q2}~,
\en
\eqn
I^{\mu\nu}_{Q2} = -i\int{d^Dpp^{\mu}p^{\nu}\over{\left(p^2-m^2+i\varepsilon
\right)\left((p+q)^2-m^2+i\varepsilon\right)}} =
\enn
\eq
{\pi}^{D\over 2}\Gamma\left(2-{D\over 2}\right)\int_0^1{dx\over{\left(m^2-x
(1-x)t\right)^{2-{D\over 2}}}}\left[q^{\mu}q^{\nu}x^2-{g^{\mu\nu}\left(m^2
-x(1-x)t\right)\over{2-D}}\right]~.
\en
Here and below $q=p_A-p_{A'}=p_{B'}-p_B~,~~t=q^2$~.
\vskip.3cm
For massless quarks the integrals (A.1)-(A.3) become the corresponding
integrals presented in~\cite{FL2}:
\eq
I_{Q2}|_{m=0} = I_2 = {\pi}^{D\over 2}(-t)^{{D\over 2}-2}{\Gamma\left(2-
{D\over2}\right)\Gamma^2\left({D\over 2}-1\right)\over{\Gamma(D-2)}}~,
\en
\eq
I^{\mu}_{Q2}|_{m=0} = I^{\mu}_2 = {q^{\mu}\over 2}I_2~,
\en
\eq
I^{\mu\nu}_{Q2}|_{m=0} = I^{\mu\nu}_2 = \left[-g^{\mu\nu}t+q^{\mu}q^{\nu}D
\right]{I_2\over{4(D-1)}}~.
\en
\vskip.3cm
To obtain the contribution of the diagram in Fig.3$\left(b\right)$ we need
the following integrals:
\eqn
I_{Q3B} = -i\int{d^Dp\over{\left(p^2-m^2+i\varepsilon\right)\left((p+q)^2-
m^2+i\varepsilon\right)\left((p_B-p)+i\varepsilon\right)}} =
\enn
\eq
-{\pi}^{D\over 2}{\Gamma\left(3-{D\over2}\right)}\int_0^1\int_0^1{d\rho_2
\over{R_{Q3}}^{3-{D\over 2}}}~,
\en
where
\eqn
d\rho_2 = dx_1dx_2\theta(1-x_1-x_2)~,
\enn
\eq
R_{Q3} = m^2(x_1+x_2)^2-x_1x_2t~.
\en
Making the substitution
\eqn
x_1 = u(1-x)~,~~~~~~x_2 = ux
\enn
we easily perform the integration over $u$ arriving to
\eq
I_{Q3B} = {\pi}^{D\over 2}{{\Gamma\left(2-{D\over2}\right)}\over 2}\int_0^1
{dx\over{\left(m^2-x(1-x)t\right)^{3-{D\over 2}}}}~.
\en
Applying the same procedure, we find
\eqn
I^{\mu}_{Q3B} = -i\int{d^Dpp^{\mu}\over{\left(p^2-m^2+i\varepsilon\right)
\left((p+q)^2-m^2+i\varepsilon\right)\left((p_B-p)+i\varepsilon\right)}} =
\enn
\eqn
-{\pi}^{D\over 2}{\Gamma\left(3-{D\over2}\right)}\int_0^1\int_0^1{d\rho_2
\over{R_{Q3}^{3-{D\over 2}}}}\left((1-x_1-x_2)p^{\mu}_B-x_2q^{\mu}\right) =
\enn
\eq
{\pi}^{D\over 2}{{\Gamma\left(2-{D\over 2}\right)}\over 2(D-3)}\int_0^1
{dx\over{\left(m^2-x(1-x)t\right)^{3-{D\over 2}}}}\left[p^{\mu}_B-(D-4)
xq^{\mu}\right]~
\en
and
\eqn
I^{\mu\nu}_{Q3B} = -i\int{d^Dpp^{\mu}p^{\nu}\over{\left(p^2-m^2+i\varepsilon
\right)\left((p+q)^2-m^2+i\varepsilon\right)\left((p_B-p)+i\varepsilon
\right)}} =
\enn
\eqn
-{\pi}^{D\over 2}{\Gamma\left(3-{D\over 2}\right)}\int_0^1\int_0^1{d\rho_2
\over{R_{Q3}^{3-{D\over 2}}}}\left[\vbox to 13.33pt{}\left((1-x_1-x_2)
p^{\mu}_B-x_2q^{\mu}\right)\right.
\enn
\eqn
\left. \times\left((1-x_1-x_2)p^{\nu}_B-x_2q^{\nu}\right)-{g^{\mu\nu}
R_{Q3}\over{4-D}}\right] =
\enn
\eqn
{{\pi}^{D\over 2}{\Gamma\left(2-{D\over 2}\right)}\over{2(D-2)(D-3)}}\int_0^1
{dx\over{\left(m^2-x(1-x)t\right)^{3-{D\over 2}}}}\left[\vbox to 10pt{}
2p^{\mu}_Bp^{\nu}_B
-(D-4)x\left(p^{\mu}_Bq^{\nu}+q^{\mu}p^{\nu}_B\right)\right.
\enn
\eq
\left. +(D-4)(D-3)x^2q_{\mu}q^{\nu}+(D-3)g^{\mu\nu}\left(m^2-x(1-x)t\right)
\right]~.
\en
In the massless case we have
\eq
I_{Q3B}|_{m=0} = I_{3B} = -2\left({D-3\over{D-4}}\right){I_2\over t}~,
\en
\eq
I^{\mu}_{Q3B}|_{m=0} = I^{\mu}_{3B} = \left(q^{\mu}-{2p_B^{\mu}\over{D-4}}
\right){I_2\over t}~,
\en
\eqn
I^{\mu\nu}_{Q3B}|_{m=0} = I^{\mu\nu}_{3B} =
\enn
\eq
\left[{1\over{(D-2)}}\left({1\over 2}g^{\mu\nu}+{p^{\mu}_Bq^{\nu}+p^{\nu}_B
q^{\mu}\over t}\right)-{q^{\mu}q^{\nu}\over{2t}}-{4p^{\mu}_Bp^{\nu}_B
\over{(D-2)(D-4)t}}\right]I_2~.
\en
\vskip.3cm
Let us now consider the case of two gluon intermediate state. To get the
contribution of the diagram in Fig. 3$\left(a\right)$, besides the integrals
$I_2$ (A.4), $I^{\mu}_2$ (A.5) and $ I^{\mu\nu}_2$ (A.6) we need
\eqn
I_{G3B} = -i\int{d^Dk\over{\left(k^2+i\varepsilon\right)\left((k+q)^2+
i\varepsilon\right)\left((p_B-k)^2-m^2+i\varepsilon\right)}} =
\enn
\eq
-{\pi}^{D\over 2}{\Gamma\left(3-{D\over2}\right)}\int_0^1\int_0^1{d\rho_2
\over{R_{G3}}^{3-{D\over 2}}}~,
\en
where
\eq
R_{G3} = m^2x_1^2-x_2(1-x_1-x_2)t~,
\en
\eqn
I^{\mu}_{G3B} = -i\int{d^Dkk^{\mu}\over{\left(k^2+i\varepsilon\right)\left(
(k+q)^2+i\varepsilon\right)\left((p_B-k)^2-m^2+i\varepsilon\right)}} =
\enn
\eq
-{\pi}^{D\over 2}{\Gamma\left(3-{D\over2}\right)}\int_0^1\int_0^1{d\rho_2
\over{R_{G3}^{3-{D\over 2}}}}\left(x_1p^{\mu}_B-x_2q^{\mu}\right)~,
\en
\eqn
I^{\mu\nu}_{G3B} = -i\int{d^Dkk^{\mu}k^{\nu}\over{\left(k^2+i
\varepsilon\right)
\left((k+q)^2+i\varepsilon\right)\left((p_B-k)^2-m^2+i\varepsilon\right)}} =
\enn
\eq
-{\pi}^{D\over 2}{\Gamma\left(3-{D\over2}\right)}\int_0^1\int_0^1{d\rho_2
\over{R_{G3}^{3-{D\over 2}}}}\left[\left(x_1p^{\mu}_B-x_2q^{\mu}\right)
\left(x_1p^{\nu}_B-x_2q^{\nu}\right)-{g^{\mu\nu}\over{4-D}}R_{G3}\right]~,
\en
and integrals $I_{G3A}$, $I^{\mu}_{G3A}$ and $I^{\mu\nu}_{G3A}$ which are
obtained from eq.s $(A.15)-(A18)$ by replacing respectively $p_B$ and $q$ with
$p_A$ and $-q$. In the zero mass limit we have
\eq
I_{G3B}|_{m=0} = I_{3B}~,~~~I^{\mu}_{G3B}|_{m=0} = I^{\mu}_{3B}~,~~~
I^{\mu\nu}_{G3B}|_{m=0} = I^{\mu\nu}_{3B}~,
\en
where $I_{3B}$, $I^{\mu}_{3B}$ and $I^{\mu\nu}_{3B}$ are given by formulas
$(A.12)-(A.14)$.
The calculation of all the integrals above listed is quite straightforward.
\vskip.3cm
We also have to consider the following more complicated integral:
\eqn
I_{G\Box} =
\enn
\eq
-i\int{d^Dk\over{\left(k^2+i\varepsilon\right)\left((k+q)^2+
i\varepsilon\right)\left((k+p_A)^2-m^2_A+i\varepsilon\right)\left((k-p_B)^2
-m^2_B+i\varepsilon\right)}}~;
\en
fortunately, we need only its asymptotic behaviour at
\eqn
s = \left(p_A+p_B\right)^2\gg -t\sim m^2_A\sim m^2_B~.
\enn
We describe its calculation with some details.
After integration over $k$ with the help of the usual Feynman parametrization
we obtain
\eq
I_{G\Box} = {\pi}^{D\over 2}{\Gamma\left(4-{D\over2}\right)}\int_0^1
\int_0^1\int_0^1{dx_1dx_2dx_3\theta(1-x_1-x_2-x_3)\over{R_{G4}}^{4-
{D\over 2}}}~,
\en
where
\eq
R_{G4} = -x_1x_2s-x_3(1-x_1-x_2-x_3)t+(x_1+x_2)\left((x_1m^2_A+x_2m^2_B
\right)~.
\en
Let us introduce new variables through relations
\eq
x_1 = zu~,~~~x_2 = (1-z)u~,~~~x_3 = x~.
\en
In terms of these variables we have
\eq
R_{G4} = -z(1-z)u^2s-x(1-x-u)t+u^2\left(zm^2_A+(1-z)m^2_B\right)~.
\en
Dividing the region of integration over $z$ into three subregions:
\eq
\int_0^1dz = \int_0^{\delta}dz+\int_{\delta}^{1-\delta}dz+\int_{1-\delta}^1
dz~,
\en
where
\eqn
1\gg \delta \gg {|t|\over s}\sim{m^2_A\over s}\sim{m^2_B\over s}~,
\enn
we can use different approximations for $R_{G4}$ in each of them:
\eq
R_{G4}~\approx~-zu^2s-x(1-x-u)t+u^2m^2_B
\en
in the first subregion,
\eq
R_{G4}~\approx~-z(1-z)u^2s-x(1-x-u)t
\en
in the second and
\eq
R_{G4}~\approx~-(1-z)u^2s-x(1-x-u)t+u^2m^2_A
\en
in the third. As $R_{G4}$ depends on $m^2_B$ in the first subregion only
and on $m^2_A$ in the third only, we can write $I_{G\Box}$ as the sum of
three terms:
\eq
I_{G\Box} \approx I_{\Box}+\triangle_B+\triangle_A~.
\en
Here $I_{\Box}$ is $I_{G\Box}$ in the massless case, $\triangle_B$ is given
by
\eqn
\triangle_B \approx {\pi}^{D\over 2}{\Gamma\left(4-{D\over2}\right)}
\int_0^1udu
\int_0^{1-u}dx\int_0^{\delta}dz\left[{1\over{\left(-zu^2s-x(1-x-u)t+u^2m^2_B
\right)^{4-{D\over2}}}}-\right.
\enn
\eqn
\left.{1\over{\left(-zu^2s-x(1-x-u)t\right)^{4-{D\over2}}}}
\right] \approx
\enn
\eqn
{{\pi}^{D\over 2}{\Gamma\left(3-{D\over2}\right)}\over {-s}}\int_0^1{du
\over u}\int_0^{1-u}dx
\enn
\eq
\times\left[{1\over{\left(-x(1-x-u)t+u^2m^2_B\right)^{3-
{D\over2}}}}-{1\over{\left(-x(1-x-u)t\right)^{3-{D\over2}}}}\right]~,
\en
and $\triangle_A$ can be obtained from $\triangle_B$ by substituting $m_B$
with $m_A$. The integral $I_{\Box}$ was calculated in ref.~\cite{FL2}:
\eq
I_{\Box} = -2(D-3)\left[\ln\left({-s\over{-t}}\right)-2\psi\left(
{D\over 2}-2\right)+\psi\left(3-{D\over2}\right)+\psi(1)\right]
{I_2\over st}~,
\en
where $\psi(z)={{\Gamma'(z)}\over{\Gamma(z)}}$ and $I_2$ is given by eq.
(A.4).
\vskip.3cm
Finally, substitution $p_B\leftrightarrow -p_{B'}=-(p_B+q)$ in eq.(A.20)
leads to the last integral we need: it can be obtained from
$I_{G\Box}$ simply by changing $s$ with $u\approx -s$.
\vskip.3cm
At last, in calculating the contribution of the diagram in
Fig. 3$\left(b\right)$ in
the case of two gluon intermediate state we meet again the integrals
$I_{G3B}$, $I^{\mu}_{G3B}$ and $I^{\mu\nu}_{G3B}$, presented in eq.s
(A.15), (A.17) and (A.18) correspondingly.

\newpage

\centerline{\bf Figure Captions}
\vskip .3 cm
\begin{description}

\item{Fig.1:}
Amplitude of the elastic scattering process $A+B\rightarrow A'+B'$ with two
particle intermediate state in the $t$-channel.
\item{Fig.2:}
Decomposition of the elastic scattering amplitude in two parts,
respectively asymptotic and non asymptotic.
\item{Fig.3:}
Contributions to the amplitude of Fig. 1, coming from the product of $(a)$
asymptotic-asymptotic parts, $(b)$ asymptotic-nonasymptotic parts, $(c)$
nonasymptotic-asymptotic parts and $(d)$ nonasymptotic-nonasymptotic parts.

\end{description}


\newpage

\begin{thebibliography}{99}

\bibitem{FKL}
V.S. Fadin, E.A. Kuraev and L.N.Lipatov, Phys. Lett. {\bf B60} (1975) 50;
E.A.Kuraev, L.N. Lipatov and V.S. Fadin, Sov. Phys. JEPT {\bf44} (1976) 443;
Sov. Phys. JEPT {\bf 45} (1977) 199.

\bibitem{LF}
L.N. Lipatov and V.S. Fadin, ZHETF Pis'ma {\bf 49} (1989) 311;
L.N. Lipatov and V.S. Fadin, Yad. Fiz. {\bf 50} (1989) 1141.

\bibitem{FL}
V.S. Fadin and L.N. Lipatov, Nucl. Phys. {\bf B} (Proc. Suppl.) {\bf 29A}
(1992) 93.

\bibitem{FL2}
V.S. Fadin and L.N. Lipatov,"Radiative Corrections to QCD Scattering
Amplitudes in a Multi-Regge Kinematics". preprint DTP/93/08, IPNO/TH 93-04,
University of Durham, 1993; submitted to Nucl. Phys. {\bf B}.

\bibitem{FF}
V.S. Fadin and R. Fiore, Phys. Lett. {\bf B294} (1992) 286.

\end{thebibliography}
\end{document}